\documentclass{sig-alt-full} 
\usepackage{times}
\usepackage{url}
\usepackage{epsf}
\usepackage{latexsym}
\usepackage{xspace}

\newcommand{\autofocus}{{\sc AutoFocus}\xspace}
\def\NM{Net\-work\-Master}
\def\AF{{\sc Auto\-Focus}}
\def\REG{Reg\-is\-try\-Mgr}
\def\REQ{Re\-quest\-Mgr}
\def\SCC{system configuration check}

\begin{document}

\conferenceinfo{ICSE'05,} {May 15--21, 2005, St. Louis, Missouri, USA.}
\CopyrightYear{2005}
\crdata{1-58113-963-2/05/0005}

\title{One Evaluation of Model-Based Testing and its Automation}

\numberofauthors{3}
\author{
\alignauthor \hspace{-.5cm}A.~Pretschner\titlenote{Corresponding
  author; {\tt pretscha@inf.ethz.ch}}\vspace{.9cm}\\
\affaddr{\hspace{-.5cm}Information~Security\\
  \hspace{-.5cm}ETH~Z\"urich\\ \hspace{-.5cm}IFW C45.2, ETH Zentrum\\ \hspace{-.5cm}8092 Z\"urich\\ \hspace{-.5cm}Switzerland}
%\email{Alexander.Pretschner@inf.ethz.ch}\\
\alignauthor \hspace{-1.8cm}W.~Prenninger\titlenote{Partially supported by the
  DFG within the priority program SoftSpez (SPP1064), project InTime.}\\ \hspace{-1.8cm}S.~Wagner\\ \hspace{-1.8cm}C.~K\"uhnel\vspace{.02cm}\\
\affaddr{\hspace{-1.8cm}Institut f\"ur Informatik\\
  \hspace{-1.8cm}TU~M\"unchen\\ \hspace{-1.8cm}Boltzmannstr. 3\\
  \hspace{-1.8cm}85748 Garching}\\ \hspace{-1.8cm}Germany\\
%\email{Wolfgang.Prenninger@in.tum.de}\\
\alignauthor \hspace{-2.5cm}M.~Baumgartner\\ \hspace{-2.5cm}B.~Sostawa\\ \hspace{-2.5cm}R.~Z\"olch\vspace{.1cm}\\
\affaddr{\hspace{-2.5cm}BMW AG, EI-20\\ \hspace{-2.5cm}Knorrstr. 147\\
\hspace{-2.5cm}80339 M\"unchen\\ \hspace{-2.5cm}Germany\\}
%\email{Martin.Baumgartner@bmw.de}\\
\alignauthor \hspace{-4cm}T.~Stauner\vspace{.9cm}\\
\affaddr{\hspace{-4.3cm}BMW~CarIT~GmbH\\ \hspace{-4.3cm}Petuelring
  116\\\hspace{-4.3cm}80809 M\"unchen\\ \hspace{-4.3cm}Germany\\}
%\email{Thomas.Stauner@bmw-carit.de}
}

\maketitle
\thispagestyle{empty}

\begin{abstract}
\noindent Model-based testing relies on behavior models
for the generation of model traces: input and expected output---test
cases---for an implementation.
We use the case study of an automotive network controller  to assess different test suites in terms of error
detection, model coverage, and implementation coverage.  Some of these
suites were generated automatically with and without models, purely at
random, and with dedicated functional test selection criteria.  Other suites were
derived manually, with and without the model at hand. Both
automatically and manually derived model-based test suites 
detected significantly more requirements errors than hand-crafted
test suites that were directly derived from the requirements. 
The number of detected
programming errors did not depend on the use of models.  Automatically
generated model-based test suites detected as many errors as
hand-crafted model-based suites with the same number of tests. A
sixfold increase in the number of model-based tests led to an 11\%
increase in detected errors.

%%% Local Variables: 
%%% mode: latex
%%% TeX-master: "main"
%%% End: 

\end{abstract}

%%A category including the fourth, optional field follows...
%\category{D.2.8}{Software Engineering}{Metrics}[complexity measures, performance measures]



%-------------------------------------------------------------------------
\section{Introduction}
\label{sec:intro}
A classical estimate relates up to 50\% of the overall development
cost to testing. Although this is likely to also include debugging
activities \cite{fagan:insp:02}, testing does and will continue to be
one of the prevalent methods in quality assurance of software systems.
It denotes a set of activities that aim at showing that a system's
intended and actual behaviors do not conform, or to increase
confidence that they do.

The intended behavior is described in specification documents that
exhibit a tendency to be incomplete, ambiguous, and sometimes
contradictory. Designing tests from such documents consequently is a
questionable undertaking.  The idea of model-based testing is to make
the intended behavior explicit, in the form of behavior models. Once
these models have been determined to accurately reflect the actual
requirements, traces of the model can serve as test cases for a
respective implementation. This approach is particularly appealing
because it is widely undisputed that in addition to the benefits
of---possibly even automated---testing, the mere activity of modeling
does help with clarifying requirements: in order to be useful,
(executable) behavior models are often so precise that they actually
form prototypes. Benefits of the latter have been acknowledged for at
least two decades \cite{boehm:prototypes:84}.

The past years have witnessed increasing research efforts on different
flavors of model-based testing.  However, we feel that the key
question has been neglected: does the approach pay off in terms of
quality and cost?  This paper
provides some answers in terms of quality. We built a
model of a network controller for modern automotive infotainment
systems to assess one representative approach to automated model-based
testing.

Throughout this paper, we use the term \emph{failure} to denote an
observable difference between actual and intended behaviors; the
reasons for the failure (incorrect state, inadequate code,
misunderstood requirements) are, without differentiation, referred to
as \emph{errors}.

\subsection{Problem} 
We address the following questions.
(1) How does the quality of model-based tests compare to traditional
hand-crafted tests? Our notion of quality covers both coverage and
number of detected failures.
(2) How does the quality of hand-crafted tests---both with and without
a model---compare to automatically generated tests, i.e., is
automation helpful? Our notion of automation relies on
characterizations of ``interesting'' test cases formalized by test
case specifications.
(3) How do model and implementation coverages relate? (4) What is the
relationship between condition/decision (C/D) coverage and failure
detection?  We do not consider cost in this paper.

\subsection{Results and Consequences} 
Our main results are summarized as follows.
(1) Tests derived without using a model detect fewer failures than
model-based tests. The number of detected \emph{programming errors} is
approximately equal, but the number of detected \emph{requirements
  errors}---those that necessitate changing the requirements
documents---is higher.
(2) Automatically generated test suites detect as many failures as
hand-crafted model-based test suites with the same number of tests. A
sixfold increase in the number of automatically generated tests leads
to 11\% additionally detected errors. None of the test suites detected
all errors. Hand-crafted model-based tests yield higher model coverage
and lower implementation coverage than the automatically generated
ones.
(3) There is a moderate positive correlation between model and
implementation C/D coverages.
(4) There is a moderate positive correlation between C/D
implementation coverage and failure detection, and a strong positive
correlation between C/D model coverage and failure detection. Higher
C/D coverage at the levels of both the model and the implementation
does not necessarily imply a higher
failure detection rate. 

Implications are threefold.
(1) In terms of failure detection, the use of models pays off.
(2) Even if entire domains can be identified where
\emph{implementation} C/D coverage strongly correlates with failure
detection, this does not necessarily mean that these positive results
carry over when the same criteria are used for automated test case
generation from \emph{models}.
(3) If the number of actually executed test cases matters, evidence
for the benefits of automated test case generation remains to be
provided.

We are aware that our findings do not necessarily generalize (see
Sec.~\ref{sec:discussion}). We think that a set of publicly accessible
medium and large-scale studies like this one will allow to draw more
general conclusions in the future.

\subsection{Experimental setup}
In a first step, we used existing requirements documents---inform\-al
message sequence charts (MSCs)---to build an executable behavior model
of the network controller. This revealed inconsistencies and omissions
in the specification documents which were updated accordingly.
They were then used
(1) by developers of a third-party software simulation of the
  cont\-roller---our system under test,
(2) by test engineers who, without the model, had to test this
  system, and
(3) by different engineers who both manually and automatically
  derived tests on the grounds of the model.

The test suites were applied to the implementation; failures were
counted and classified. The model itself, as ``ultimate reference'',
was not included in the requirements documents. This explains why
  there are requirements errors at all: the updated specification MSCs
  did not capture all the implementation behaviors that, later on,
  exhibited mismatches with the model's behavior.

\subsection{Contribution} 
We are not aware of studies that systematically compare automatically
generated test suites to hand crafted ones. We are also not aware of
real-world studies that try to precisely measure the benefits of using
explicit models for testing as opposed to not using them. We see our
major contribution in providing numbers that indicate the usefulness
of explicit behavior models in testing, and in stimulating the
discussion on the usefulness of automation and the use of structural
criteria in model-based testing.

\subsection{Overview} 
Sec.~\ref{sec:MBT} defines our notion of model-based testing in
general, the modeling tool we used, and the technology of test case
generation.  Sec.~\ref{sec:most} gives an overview of the network
controller.  Sec.~\ref{sec:tests} presents different test suites and
their performance.  Sec.~\ref{sec:discussion} discusses the findings
of our case study, Sec.~\ref{sec:relwork} describes related work,
and Sec.~\ref{sec:concl} concludes.

%%% Local Variables: 
%%% mode: latex
%%% TeX-master: "main"
%%% End: 

\section{Model-Based Testing}
\label{sec:MBT}

This section provides a description of model-based testing in general,
the CASE tool \autofocus, and a sketch of the generation of test cases
from \autofocus models.

\subsection{Basics}
\label{subsec:basicsMBT}
The general idea of model-based testing (of deterministic systems) is
as follows.
An explicit behavior model encodes the intended behavior of an
implementation called \emph{system under test}, or SUT. Modeling
languages include statecharts, Petri nets, the UML-RT, or ordinary
code.  Traces of the model are selected, and these traces constitute
test cases for the SUT: input and expected output.

The model must be more abstract than the SUT. In general, abstraction
can be achieved in two different ways:
(1) by means of encapsulation: macro-like structures as found
in compilers, library calls, the MDA, or J2EE, or
(2) by deliberately omitting details and losing information such as
timing behavior.
Now, if the model was not more abstract than the SUT in the second
sense, then the efforts of validating the model would exactly match
the efforts of validating the SUT. (We use the term validation when an
artifact is compared to often implicit, informal
requirements.)  

While the use of abstraction in model-based testing appears
methodically indispensable, and, for the sake of intellectual mastery,
desirable, it incurs a cost: details that are not encoded in the model
obviously cannot be tested on the grounds of this model. In addition,
it entails the obligation of bridging the different levels of
abstraction between model and SUT: input to the model, as given by a
test case, is concretized before it is fed to the SUT.
The output of the latter is abstracted
before it is compared to the output of the model as defined by the
test case. The hope is that one can split the inherent complexity of a
system into
an abstract model, and
driver components that perform concretizations and abstractions.
The granularity of the comparison between the system's and the model's
output depends on the desired precision of the test process: as an
extreme case, each output can be abstracted into whether or not an
exception was thrown.  In some situations, this may be meaningful
enough to initiate further actions.

In most cases, one needs selection criteria on the set of all traces
of the model. We call them \emph{test case specifications}. These are
intensional: rather than specifying each test case on its own, one
specifies a characteristics and has some manual or automatic generator
derive test cases that exhibit the characteristics. Examples include
coverage criteria, probability distributions, or the definition of a
state of the model one considers interesting. They can also be given
by functional requirements in the form of restricted environment
models that make sure the model of the SUT can only perform certain
steps \cite{PhilippsPSAKS03}. This also includes fault models. In this
sense, test case specifications can be structural, stochastic, or
functional.

To summarize the procedure in the present study, we built a model of
the network controller and a rudimentary environment model for the
nodes in the network. As far as model-based tests are concerned, this
model together with the test case specifications of
Sec.~\ref{subsec:tcspecs} was used to derive a set of model traces, or
runs. There are no explicit fault models; these are implicitly
represented in the test case specifications. By projecting a trace
onto the behavior of the controller, we get a test case for its implementations:
input and expected output. With suitable concretizations and
abstractions, we stub the actual nodes by the information contained in
the test case. The network itself as well as the controller were not
stubbed.

\subsection{AutoFocus}  
\label{subsec:AF}

\begin{figure}[t]
\begin{center}
  \includegraphics[width=.36\textwidth]{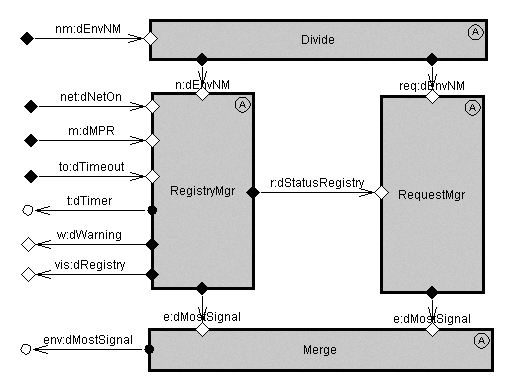}
  \caption{SSD of the MOST \NM}
  \label{fig:SSD}

\end{center}
\end{figure}

We use the CASE tool \autofocus{} \cite{hse:fme97}
for modeling the network controller.
The core items of \autofocus{} specifications are components. A
component is an independent computational unit that communicates with
its environment via so called ports. Ports are typed.  Two or more
components can be linked by connecting their ports with directed
channels. In this way, component networks evolve which are described
by \emph{system structure diagrams} (SSDs). As an example,
Fig.~\ref{fig:SSD} shows the structure of the network controller. It
is explained in Sec.~\ref{subsec:NMmodel}.

SSDs are hierarchical. This means that each component can recursively be
described as a set of communicating subcomponents.  Atomic components
are components which are not further refined. For these components a
behavior must be defined. This is achieved by means of extended finite
state machines (EFSMs). 
Fig.~\ref{fig:STD}, also described in Sec.~\ref{subsec:NMmodel},
depicts the EFSM of a central component of the network controller.

An EFSM consists of a set of control states (bubbles), transitions
(arrows), and is associated with local variables. The local variables
form the component's data state. Each transition is defined by its
source and destination control states, a guard with conditions on
input and the current data state, and an assignment for local
variables and output ports.  Transitions can fire if the condition on
the current data state holds and the actual input matches the input
conditions.  Assignments modify local variables. After execution of
the transition, the local variables are set accordingly, and the
output ports are bound to the values computed in the output statements.
These values are then copied to the input ports that are connected by
channels.  Guards and assignments are defined in a Gofer-like
functional language that allows for the definition of possibly
recursive data types and functions.

\autofocus{} components are timed by a common global clock: they all
perform their computations simultaneously. Each clock cycle consists
of two steps: first each component reads the values on its input ports
and computes new values for local variables and output ports. New
values are copied to the output ports where they can be accessed
immediately via the input ports of connected components.
In addition to the simple
time-synchronous execution and communication semantics of \AF, it is
the use of a functional language that makes \AF\ models amenable to
test case generation.

\subsection{Test Case Generation}
\label{subsec:testgen}

Test case generation is performed by translating the model into a
Constraint Logic Programming (CLP) language, and adding the test case
specification---a full-fledged environment model, or sets of
constraints.
Execution of this CLP program then successively enumerates all traces
of the model (and ``guesses'' all possible input values). 
In fact, the model is executed symbolically: rather than enumerating
single traces---input, output, local data of all compo\-nents---of the
model, we work with sets of values in each step instead.  States are
not visited multiple times which is why in each step, the currently
visited set of states is only taken into consideration if it is not a
specialization of a previously visited set of states.
We omit details of the translation and state storage here and refer to
earlier work \cite{psak:MBTinhouse:04}.

Even with test case specifications and state storage, the number of
computed test cases that satisfy a test case specification may be too
large. In this case, one has to add further constraints, i.e., test
case specifications, or pick some tests at random.

%%% Local Variables: 
%%% mode: latex
%%% TeX-master: "main"
%%% End: 

%\newpage
\section{The MOST \NM}
\label{sec:most}

\emph{MOST (Media Oriented Systems Transport)} is an infotainment
network tailored to the automotive domain. Its public specification
\cite{most-spec} is maintained by the MOST cooperation that includes
major automotive companies. This public specification does not contain
the informal sequence diagrams that we used in our study. 
MOST is a
ring topology that supports synchronous and asynchronous communication
at up to 24.8 Mbps.
Various devices, such as a CD changer or a navigation system, are
connected in order to provide MOST applications to the
user.  These applications are represented by \emph{function blocks}
that reside in MOST devices. Examples of a function block include 
\emph{CDPlayer} and the special function block \emph{NetBlock}. This
function block is available in every device and can be used to get
information about the other function blocks. Each function block
provides several functions that can be used by other function blocks.
For instance, a \emph{CDPlayer} can be asked to \emph{start}, \emph{stop},
etc. All function blocks and functions are addressed by standardized
identifiers.

The network exhibits three central master function blocks, one of
which is the \NM{} (NM), the subject of our study.
It is responsible for ensuring consistency of the various function
blocks, for providing a lookup service, and for assigning logical
addresses.

\subsection{Model of the NetworkMaster}
\label{subsec:NMmodel}
Fig.~\ref{fig:SSD} depicts the functional decomposition of the NM\ 
into \AF\ components. The NM provides two basic services.  The first
is to set up and maintain the \emph{central registry}. The central
registry contains all function blocks and their associated network
addresses currently available in the network. This service is modeled
by component \texttt{\REG}. The second service is to provide a lookup
service from function blocks to network addresses.  This task is
modeled by component \texttt{\REQ}.  Components \texttt{Divide} and
\texttt{Merge} are needed for technical reasons; they distribute
incoming and merge outgoing signals.

\begin{figure}[t]
  \includegraphics[width=.46\textwidth]{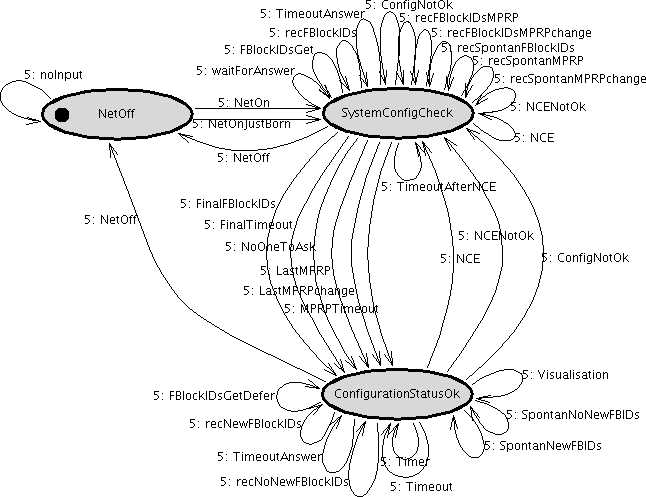}
  \caption{RegistryManager's Behavior}
  \label{fig:STD}
\end{figure}

Fig.~\ref{fig:STD} depicts the EFSM of component \texttt{\REG} which
is the most complex in the model. For the sake of simplicity, we do
not provide any guards and actions on transitions here. The
component's data space is partitioned into three control states
(bubbles): \texttt{Net\-Off} models the state when the NM is switched
off; in state \texttt{System\-Config\-Check} the NM performs a \SCC,
i.e., it sets up or checks the central registry; and in state
\texttt{Con\-fig\-ura\-tion\-Status\-Ok} the MOST network is in normal
operation, i.e., the nodes in the network are allowed to communicate
freely.

Including the environment model, the model consists of 17 components
with 100 channels and 138 ports, 12 EFSMs, 16 distinct control states
(bubbles), 16 local variables, and 104 transitions. 34 data types were
defined by means of 80 constructors.  The number of defined
functions---used in guards and assignments---is 141. The model's
  complexity lies in these functions and in the transition guards. The part of
  the implementation that the model roughly corresponds to, amounts to 12,300
  lines of C code, without comments.

Five general abstraction principles were applied in the model.
\begin{enumerate}
\item In terms of \emph{functional abstraction}, we focused on the
  main functionality of the NM, namely setting up and maintaining the
  registry, and providing the lookup service. We omitted node
  monitoring which checks from time to time whether or not all nodes
  in the ring are alive.
\item In terms of \emph{data abstraction}, we reduced data complexity
  in the model, e.g., by narrowing the set of MOST signals to those
  which are relevant for the NM behavior, and by building equivalence
  classes on error codes which the NM treats identically.
\item In terms of \emph{communication abstraction}, we merged
  consecutive signals that concern the same transaction in actual
  hardware into one signal.
\item In terms of \emph{temporal abstraction}, we abstracted from
  physical time. For instance, the timeout that indicates expiration
  of the time interval the NM should wait for the response of a node
  is abstracted by introducing two symbolic events: one for starting
  the timer, and a nondeterministically occurring one for expiration
  of the timer. This nondeterministic event is raised outside
    of the NM model which hence remains deterministic.
\item Finally, in terms of \emph{structural abstraction}, the nodes in
  the environment of the NM are not represented as \AF\ components,
  but instead by recursive data structures manipulated by one
  dedicated environment component. This enables us to parameterize the
  model in order to deal with a variable number of nodes in the
  network.
\end{enumerate}

\subsection{Implementation} 
The SUT is a beta software simulation of the MOST NM that is connected
to an actual network.  The network controller is intended to be built
by different suppliers, not the automotive OEM who, nonetheless, needs
a software-in-the-loop simulation for integration tasks with other
devices. The NM simulation was built by an external third party.
Roughly, the interface of the SUT is identical to that of hardware
NMs.

In order to make the abstract test cases---model traces---applicable to the SUT, we
wrote a compiler that translates them into 4CS (\url{www.4cs.de}) test
programs. 4CS provides a test infrastructure for MOST networks.
So-called optolyzers were used to stub actual nodes: these are freely
programmable nodes in the MOST network. Via 4CS, we programmed them to behave like  a corresponding test case. In
this way, we can stimulate the SUT. In the 4CS programs, the SUT's
output is compared to the intended output as given by the test
case. At the end of each test case, the central registry of the SUT
was downloaded and compared to the corresponding registry of the model
which is also encoded in the test cases.

We omit details of the instantiation of the general scheme of
Sec.~\ref{subsec:basicsMBT} with driver components responsible for
input concretization and output abstraction.
For instance, in terms of data abstraction, one arbitrary
representative of an equivalence class of error codes---sent to the
SUT---was chosen in order to instantiate signals. 
In terms of temporal abstraction, the expiration of a timer was
instantiated by a wait statement containing the actual physical
duration.
Conversely, output of the model is converted into an executable
verification statement. For example, if an output signal contains a
list of items as parameter, a corresponding verification statement is
created which checks if the actual list in the implementation's output
is a permutation of the expected list in the model's output: the model
is deliberately over-specified.

%%% Local Variables: 
%%% mode: latex
%%% TeX-master: "main"
%%% End: 

\section{Tests}
\label{sec:tests}

This section describes the general procedure of testing the NM,
different test suites, and observations.

\subsection{Overview}
\label{subsec:testsProcedure}
Once the model 
had been built, we derived different test suites
(Sec.~\ref{subsec:testsuiteCoverage}). Except for hand-crafted test
cases, these consist of abstract sequences of input and expected
output. We turned them into executable test cases as described in
Sec.~\ref{sec:most}. Tests built without a model were manually lifted
to the more abstract level of the model. Doing so allows us to (1)
apply all test cases to the implementation via the 4CS compiler, (1a)
check for conformance with the model, and (1b) measure coverage at the
level of the implementation. In addition, we (2) applied the input
part of each test case to the model and measured coverage at the level
of the model.  \emph{Model coverage} is defined by means of coverage
on Java (simulation) code that was generated from the model.
\emph{Implementation coverage}, on the other hand, was measured on the
C code of the SUT. For the sake of comparability, we excluded those C
functions that, as a consequence of abstraction, do not have
counterparts in the model. However, some of the abstracted behavior is
scattered over the C code, and we did not remove these parts for
measurements.

Our coverage criterion is based on the control-flow of a program.
\emph{Condition/Decision (C/D) coverage} measures the number of
different evaluations (a) of each atomic condition in a composed
condition plus (b) the outcome of the decision. 100\% coverage
requires that each atomic condition be evaluated at least once to both
true and false, plus the requirement that the decision takes both
possible outcomes.

In addition to coverage measurements, we recorded differences between
the behaviors of model and implementation, and grouped these failures
into 26 classes. Since the elements of a class exhibit a similar
erroneous behavior, we conjecture that the elements of each class
correspond to the same fault, i.e., the same cause of the deviation in
behaviors. Since the SUT was built by an external third party, we
could not verify this conjecture. Consistent with the terminology
introduced in Sec.~\ref{sec:intro}, we use the terms ``failure class''
and ``error'' interchangeably. When talking about numbers of detected
errors, we always mean \emph{distinct} errors.

Different test suites were applied in order to assess
(1) the use of models vs.\ hand-crafted tests,
(2) the automation of test case generation with models, and
(3) the use of explicit test case specifications.  We also provide a
comparison with randomly generated tests.

\subsection{Test Suites}
\label{subsec:testsuiteCoverage}
This section describes the seven different test suites that we
compared, and explains to what end we designed them.  The length of
all test cases varies between 8 and 25 steps (our test case generator handles test cases of arbitrary finite lengths, but for the sake of human analysis within this study, we restricted ourselves to rather short sequences). To all test cases, a
postambule of 3-12 steps is automatically added that is needed to
judge the internal state of the SUT (registry download). We are
concerned with black-box testing an NM implementation, i.e., we do not
directly access its internal state. However, because it is a software
simulation, we can easily measure code coverage.

\begin{table}[t]
\centering
\caption{\label{tab:tsuites}Test suites}
{
\begin{tabular}{|c||c|c|c|}
\hline
suite & automation & model & TC specs\\ \hline\hline
$\mathcal{A}$ & manual & yes & yes\\
$\mathcal{B}$ & auto   & yes & yes\\
$\mathcal{C}$ & auto   & yes & no\\
$\mathcal{D}$ & auto   & no  & n/a\\
$\mathcal{E}$ & manual & no  & n/a\\
$\mathcal{F}$ & manual & no & n/a\\
$\mathcal{G}$ & manual & no  & n/a\\ \hline
\end{tabular}
}
\end{table}

We investigated the following test suites. The exact number of tests in suites  $\{\mathcal{B}, \mathcal{C}, \mathcal{D}\}$ is given in Sec.~\ref{subsubsec-errordetection}.
 
\begin{itemize}
\item [$\mathcal{A}$] A test suite that was \emph{manually} created by
means of \emph{interactively simulating the model}: $|\mathcal{A}|=105$ test cases.
\item [$\mathcal{B}$] Test suites that were generated \emph{automatically}, on the grounds of the model, \emph{by taking into account the functional test case specifications} of
Sec.~\ref{subsec:tcspecs}. Tests were generated at random, with
additional constraints that reflect the test case specifications.  The
number of test cases in each suite varies between 40 and 1000. We
refer to these test suites as ``automatically generated''.
\item [$\mathcal{C}$] Test suites that were generated at random, 
\emph{automatically} on the grounds of the model, \emph{without taking
  into account any functional test case specifications}.
\item [$\mathcal{D}$] Test suites that were
\emph{randomly} generated, \emph{without referring to the
  model}. Sec.~\ref{subsubsec:generation} explains how the expected
output part was derived.
\item [$\mathcal{E}$] A \emph{manually} derived test suite that
represents the \emph{original requirements message sequence charts}
(MSCs).  This test suite contains $|\mathcal{E}|=43$ test cases.
\item [$\mathcal{F}$] A \emph{manually} derived test suite that, in
addition to the original requirements MSCs, contains some further
MSCs. These are a result of \emph{clarifying the requirements} by
means of the model. The test suite itself was derived without the
model. This test suite contains $|\mathcal{F}|=50$ test cases.
\item [$\mathcal{G}$] A test suite that
was \emph{manually developed} with traditional techniques, i.e.,
without a model: 61 
test cases.
\end{itemize}

All these test suites are summarized in Tab.~\ref{tab:tsuites}. The
difference between test suites $\{\mathcal{E},\mathcal{F}\}$ and $\mathcal{G}$ is that $\{\mathcal{E},\mathcal{F}\}$ directly
stem from requirements documents only whereas $\mathcal{G}$ stems from test
documents (which, of course, rely on requirements themselves). The
difference between  $\mathcal{A}$ and $\mathcal{F}$ is similar: $\mathcal{F}$ is a direct
result of requirements engineering activities, and $\mathcal{A}$ results from
testing activities.

\subsubsection{Functional Test Case Specifications (suite $\mathcal{B}$)}
\label{subsec:tcspecs}
We defined functional test case specifications in order to specify
sets of test cases to be generated for suite $\mathcal{B}$. Each test
specification is related to one functionality of the NM, or to a part
of the behavior it exhibits in special situations.  We identified
seven classes of functional test case specifications that we state
informally.

\begin{itemize}
\item [\bf TS1] Does the NM start up the network to normal operation
  if all devices in the environment answer correctly?
  
\item [\bf TS2] How does the NM react to central registry queries?
  
\item [\bf TS3] How does the NM react if nodes don't answer?
  
\item [\bf TS4] Does the NM recognize all situations when it must
  reset the MOST network?
  
\item [\bf TS5] Does the NM register signals that occur spontaneously?
  
\item [\bf TS6] Does the NM reconfigure the network correctly if one
  node jumps in or out of the network?
  
\item [\bf TS7] Does the NM reconfigure the network correctly if a
  node jumps in or out of the network more than once ?
\end{itemize}

We implemented and refined TS1--TS7 into 33 test
case specifications by stipulating that specific signals must or must
not occur in a certain ordering or frequency in traces of the NM
model.

\subsubsection{Generation}
\label{subsubsec:generation}
Generation of test cases was done as follows. For suite $\mathcal{B}$, we
translated the specifications of Sec.~\ref{subsec:tcspecs}
into constraints, and added them to the CLP translation of the model
(Sec.~\ref{subsec:testgen}).  Each of the 33 refined test case
specifications basically consists of a conjunction of combinations of
those constraints that correspond to the specifications TS1-TS7. The
resulting CLP program was executed for test cases of a length of up to
25 steps. Computation was stopped after a given amount of time, or, as
a consequence of state storage and test case specification, when there
were no more test cases to enumerate.  For each of the 33
specifications, this yielded suites that satisfy them.  During
test case generation, choosing transitions and input signals was
performed at random. In order to mitigate the problems that are a
result of the depth first search we use, we generated tests with
different seeds for the random number generator: for each test case
specification, fifteen test suites with different seeds were computed.
Out of each of the fifteen suites, a few tests were selected at random. We hence
generated test suites that were randomly chosen from all those test
cases that satisfy the test case specifications. 

Suite $\mathcal{C}$ was generated in a similar manner, but without any functional test case
specifications. Suite $\mathcal{D}$ was derived by randomly generating input
signals that obeyed some sanity constraints (e.g., switch on the
device at the beginning of a test case) but did not take into account
any logics whatsoever. In order to get the expected output part of a
test case, we applied the randomly generated input to the model and
recorded its output.

\begin{figure}[t]
  \centering
  \includegraphics[width=.42\textwidth]{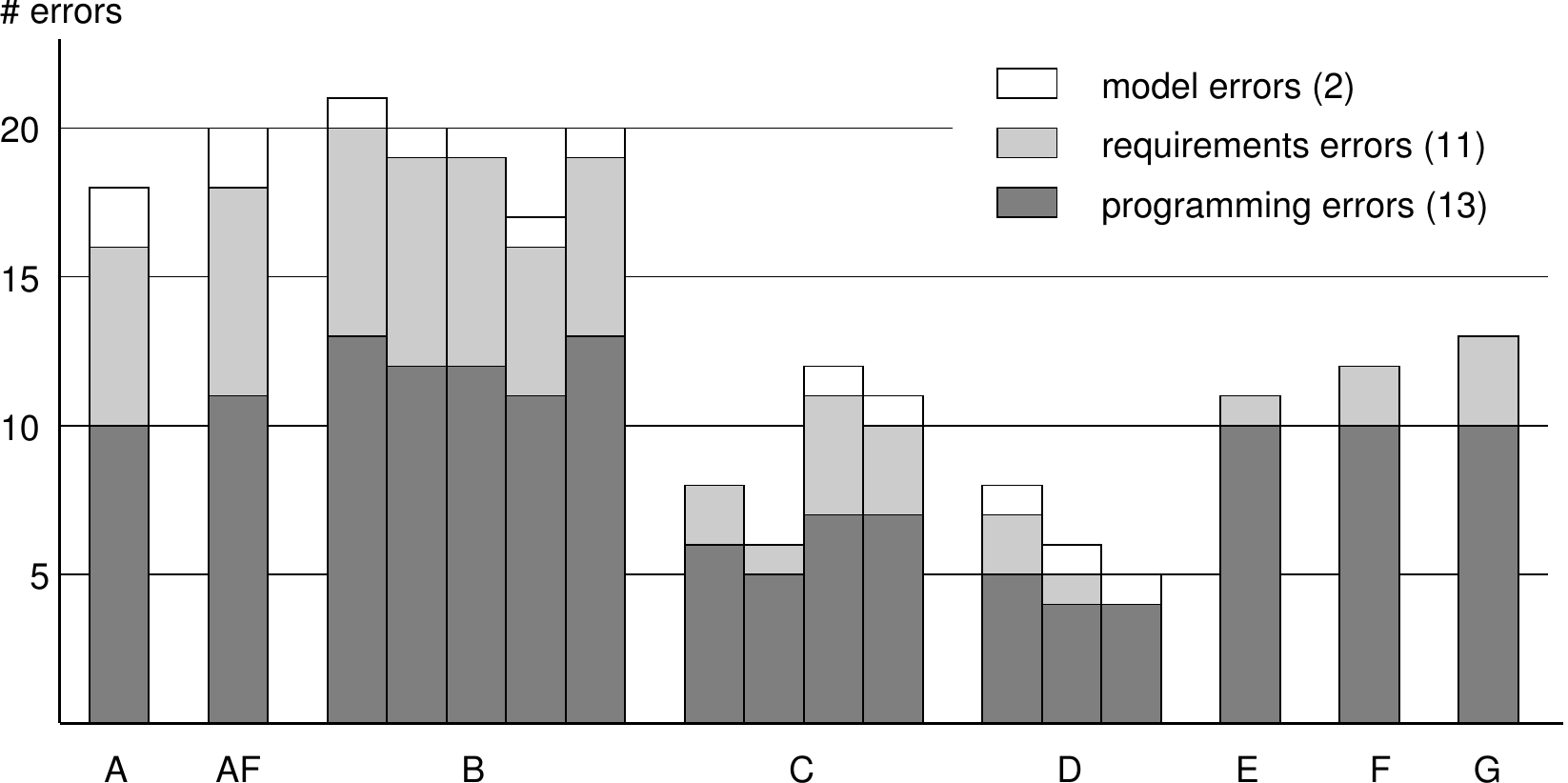}
  \caption{Detected errors}
  \label{fig:kinds-of-errors}
\end{figure}

Hand-crafted test cases were, with the exception of test suite $\mathcal{A}$,
conceived without the model. The expected output parts were derived by
applying the respective input to the model. Using the model for manual
derivation of tests here means that knowledge of the model and its
structure was an essential part of the process of designing the test
suite.

\subsubsection{Rationale} 
As mentioned earlier, we address the following questions.
\begin{enumerate}
\item \emph{Does the use of models yield better tests?} This is covered by contrasting suites $\{\mathcal{D},\mathcal{E},\mathcal{G}\}$ to $\{\mathcal{A},\mathcal{B},\mathcal{C}\}$.
The intention behind test suites $\mathcal{E}$ and
$\mathcal{F}$ is to give some comparative numbers when the mere execution of
documented requirements scenarios is considered.
\item \emph{Does automation justify the effort?}  This is covered by contrasting $\{\mathcal{B},\mathcal{C}\}$ to $\mathcal{A}$.  Test suite $\mathcal{C}$ in
comparison with $\mathcal{B}$ is used to assess the concept of test case
specifications when compared to purely random testing.
\end{enumerate}

\subsection{Observations and Interpretations}
\label{sec:observations}
This section describes our findings in terms of error detection, model
coverage, and implementation coverage.

\subsubsection{Error detection}
\label{subsubsec-errordetection}
26 errors were found during the testing phase, in addition to 3 major
inconsistencies, 7 omissions, and 20 ambiguities that were found in
the specification documents while the model was built.  Two of the 26
errors are errors in the model, a consequence of mistaken
requirements. There are 13 programming errors, and 11 requirements
errors. The latter are defined by the fact that their removal involved
changing the user requirements specifications

(recall that these did not include the model itself as ``ultimate
reference'': naturally, even the updated requirements MSCs, $\mathcal{F}$,
contained omissions and ambiguities). Changing requirements
specifications wasn't necessary for programming errors. The difference
between the two classes obviously also is important in terms of who is
responsible for the removal, the OEM or the supplier. Out of the 24
errors in the implementation, 15 were considered severe by the domain
experts, and 9 were considered non-severe. 
Severity means that their occurrence at runtime would jeopardize a
subsequent correct functioning of the entire system.  Our definition
of requirements errors inherits from the notion of design errors
coined by Boehm et al. \cite{boehm:autoaid:75}.

\begin{figure}[t]
  \includegraphics[width=.47\textwidth]{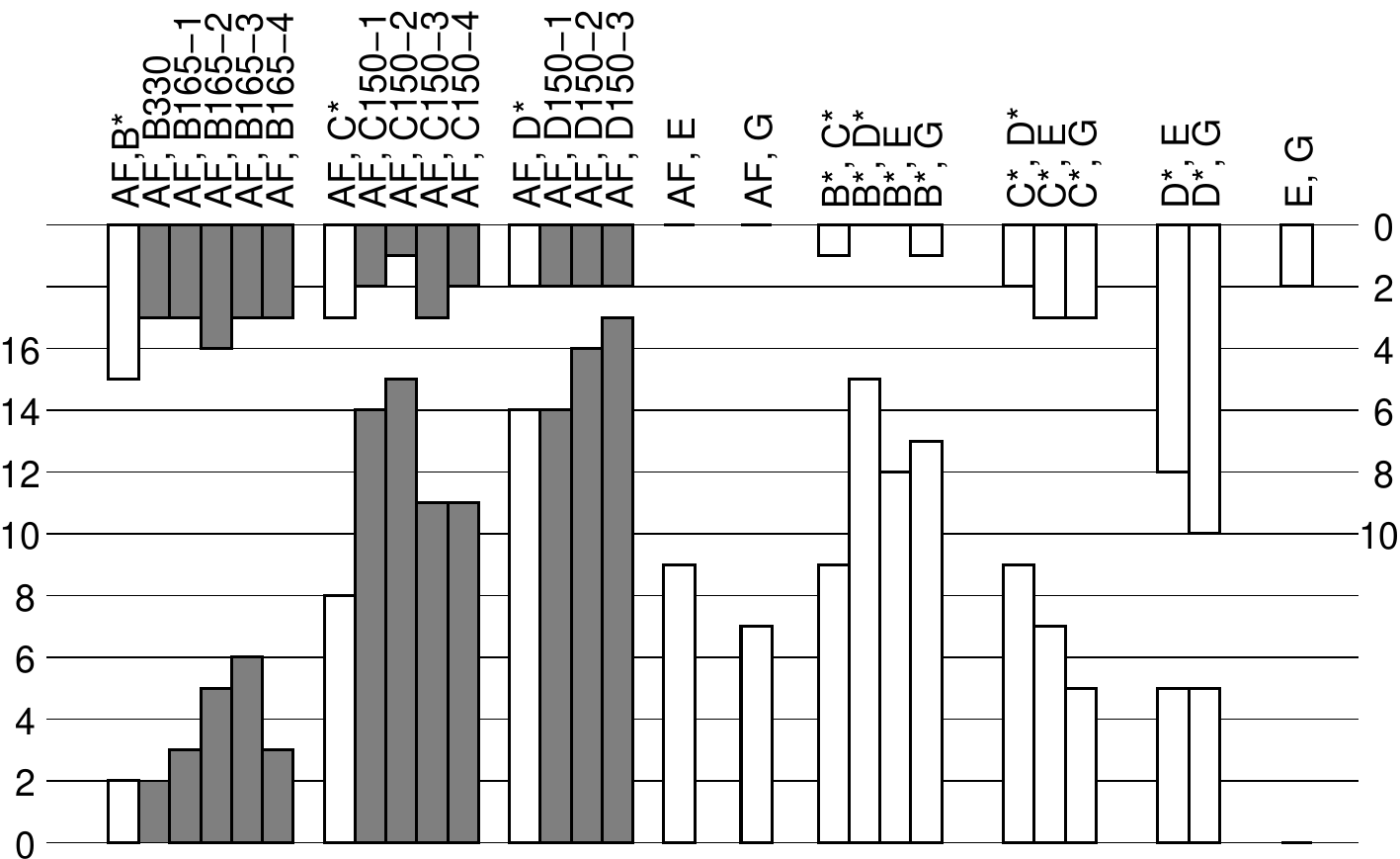}
  \caption{Differences in detected errors}
  \label{fig:errors-diff}
\end{figure}

Because we could not automatically assign a failure to its class
(error, cf. Sec.~\ref{subsec:testsProcedure}), we had to manually
check the results of running the test cases. This restricts the number
of tests. In terms of suites $\mathcal{B}$, we picked 4 times 5 tests and once 10
tests for each of the 33 refined test case specifications, which adds
to 4*165+330=990 tests. For suites $\mathcal{C}$, we picked 4 times 150 tests, and
3 times 150 tests for suites $\mathcal{D}$.  Fig.~\ref{fig:kinds-of-errors} shows
the errors (classes of failures) that were detected with different
test suites; the first bar for suite $\mathcal{B}$ is the one that consists of 330
tests. The $\mathcal{AF}$ bar represents the test suite that consists of $\mathcal{A}\cup\mathcal{F}$; these two together seemed a natural reference
candidate for assessing automated tests. Fig.~\ref{fig:errors-diff}
shows the differences in terms of numbers of detected errors. Let $\varepsilon(\mathcal{X})$ denote the errors detecetd by suite $\mathcal{X}$. For a
pair $\mathcal{X},\mathcal{Y}$, the lower bar denotes the
number of errors  detected by $\mathcal{X}$ but not by $\mathcal{Y}$: $|\varepsilon(\mathcal{X})\backslash\varepsilon(\mathcal{Y})|$. The upper bar shows the inverse:
 $|\varepsilon(\mathcal{Y})\backslash\varepsilon(\mathcal{X})|$. An asterisk, $\star$, refers to a
cumulated test suite ($\mathcal{B}, \mathcal{C}$, or $\mathcal{D}$).  In terms of suites $\mathcal{B}$, $\mathcal{C}$, and $\mathcal{D}$,
grey bars denote suites that were not cumulated. For these
suites, the number of test cases and the index of the suite are also
given. The latter corresponds to the ordering of test suites in
Fig.~\ref{fig:kinds-of-errors}. For instance, $\mathcal{B}330$ denotes test
suite $\mathcal{B}$ consisting of 330 tests, and $\mathcal{C}$150-3 denotes the third test
suite $\mathcal{C}$ consisting of 150 tests. As an example, the leftmost bars in
the figure show that $\mathcal{AF}$ detected 2 errors that the cumulated suite $\mathcal{B}^\star$
did not detect, and conversely, that $\mathcal{B}^\star$ detected 5
errors that $\mathcal{AF}$ did not detect.

The major observation is that model-based and hand-crafted tests both
detect approximately the same number of programming errors.
Requirements errors are predominantly detected by model-based tests.
This is because building the model involved a thorough review of the
requirements documents, and these are directly reflected in the model.
None of the test suites detected all 26 errors, and there is no
correlation between test suites and the severity status of the
respective detected errors (figure not shown).

Suite $\mathcal{A}$ (105 tests; 18 errors) detects slightly fewer errors than $\mathcal{AF}$ (148 tests; 20 errors). The two errors detected by $\mathcal{F}$
but not by $\mathcal{A}$ were simply forgotten; they ``should have been found''.
20 is approximately the same number as the number detected by each
single suite in $\mathcal{B}^\star$. The cumulated suite $\mathcal{B}^\star$ detects 23 errors. Note that
the latter consists of 990 tests while $\mathcal{AF}$ consists of 148 tests, and
one might well argue that $\mathcal{A}$ plus the two inattentively ``forgotten''
tests---hence 107 tests---makes for a fairer comparison than the
entirety of $\mathcal{AF}$.  The errors detected by $\mathcal{B}^\star$ but not $\mathcal{AF}$ correspond to
situations that appear abstruse to a human which makes us believe a
human tester would not have found them---that they were detected by
automatically generated suites is a consequence of the randomness
involved. These situations were judged unlikely by the domain experts.

Randomly generated model-based tests (suites $\mathcal{C}$, cumulated: 15 errors)
detect roughly as many errors as manually designed tests ($\mathcal{E}$-$\mathcal{G}$). The
latter detect more programming errors, and almost the same number of
requirements errors. With a few exceptions---again ``abstruse''
situations---errors detected by $\mathcal{C}^\star$ are also detected by $\mathcal{AF}$ and $\mathcal{B}^\star$: traces that are executed with a high probability.

Suites $\mathcal{D}$ (cumulated: 8 errors) exhibit the smallest number of detected
errors. All of them are also detected by $\mathcal{B}^\star$; two errors not detected by
$\mathcal{AF}$ correspond to traces that, once more, appear abstruse to a human
because of the involved randomness.  The use of functional test case
specifications hence ensures that respective tests perform better than
purely randomly generated tests.

\begin{figure}[t]
  \includegraphics[width=.46\textwidth]{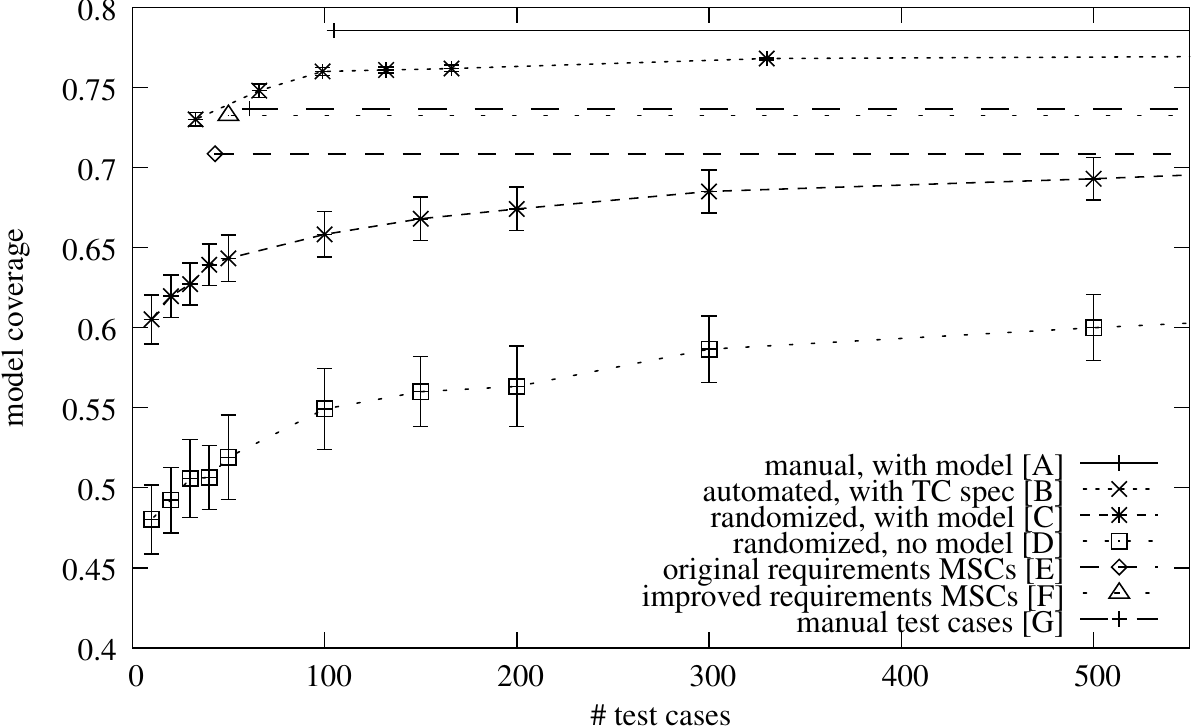}   
  \caption{Model coverage}
  \label{fig:cumulated-coverage}
\end{figure}

\subsubsection{Model Coverage}
\label{subsubsec-cov}

The model contains 1722 C/D
evaluations 
in transition guards and functional programs used by component
\texttt{Reg\-istryManager}. The implementation contains 916 C/D
evaluations.
Fig.~\ref{fig:cumulated-coverage} shows C/D coverage at the level of
the Java simulation code generated from the model.

For test suites with varying numbers of test cases, we display the
mean that was computed from 25 experiments, i.e., 25 times a choice of
$n$ test cases out of original sets that range from 6,000 to 10,000
tests.  The error bars denote the 98\% confidence interval for the
mean under the assumption that the data is approximately normally
distributed.  For the sake of graphical representation, we do not
display any numbers for more than 550 test cases.

Coverage does not exceed 79\%.  The reason is the handling of pattern
matching in the generated Java code with trivially true conditional
statements.
Except for the test cases that have been generated without a model,
the 98\% confidence intervals for the given means are rather small.
This implies a likelihood that the displayed trends are not subject to
random influences. 

$\mathcal{A}$ yields the highest coverage which is unmatched by the
second best suite $\mathcal{B}$. That $\mathcal{A}$ yields such a high coverage is explained
by the fact that the same person built the model and the test case
specifications of Sec.~\ref{subsec:tcspecs}.  This person intuitively
tried to match the structure of the model.  In our case study,
automation could hence not match the coverage of manually generated
model-based tests. $\mathcal{A}$ does not include all covered C/Ds of suites $\mathcal{B}$ to
$\mathcal{G}$: even though the absolute coverage of $\mathcal{A}$ is the highest, it turns out
that the latter yield up to 14 additional evaluations of atomic
conditions. It also turned out that generated tests covered more
possible input signals, a result of randomization. Manually derived
test cases included some special cases that the randomly generated
tests did not cover.

Suites $\mathcal{F}$ and $\mathcal{G}$ are the next best suites; this is explained by the fact
that the improved requirements documents contain some ``essential''
runs of the model. Suite $\mathcal{C}$, i.e, randomly generated model-based tests,
match the coverage of $\mathcal{F}$ at about 500 test cases.

\begin{figure}[t]

  \includegraphics[width=.46\textwidth]{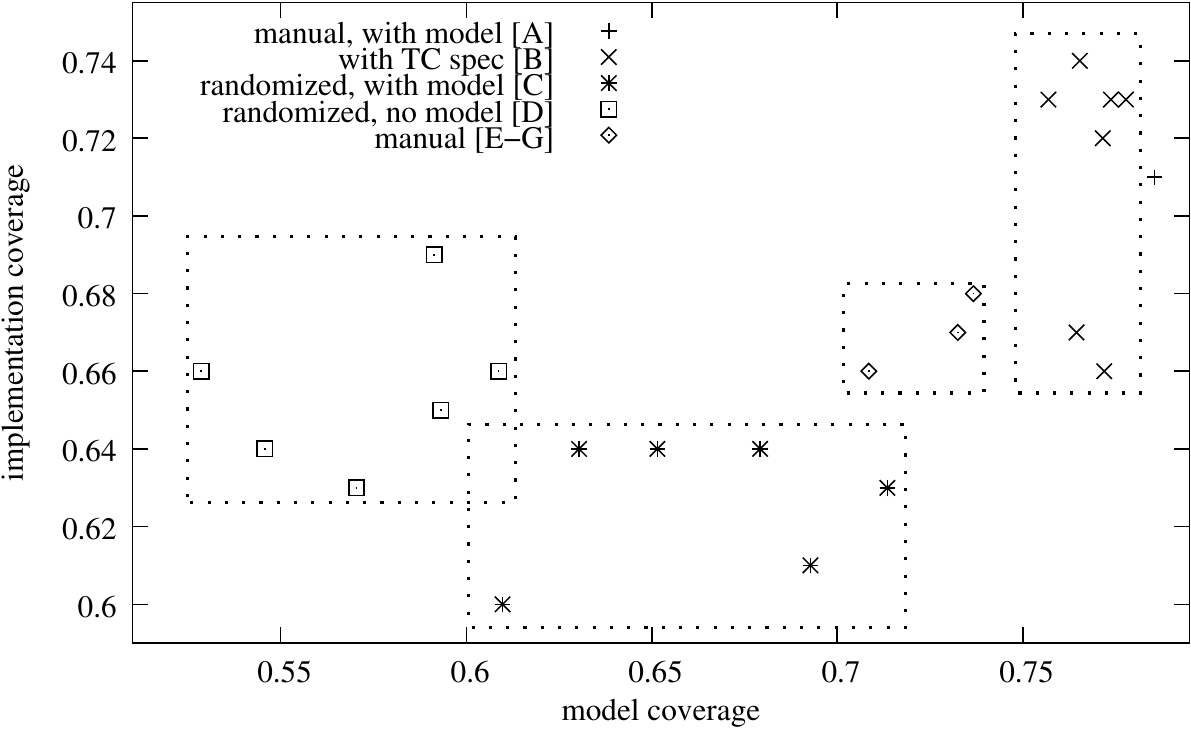}
  \caption{Coverages}
  \label{fig:cov-vs-cov}
\end{figure}

The comparison of test suites $\{\mathcal{C},\mathcal{D}\}$ and $\mathcal{B}$ shows that 
the use of functional test case specifications leads to higher
coverage with fewer test cases. This comes as no surprise since test
case specifications ``slice'' the model. If test cases are generated
for each ``slice''---which correspond to different structural
elements, or rare special conditions, in the model---then there is an
increased likelihood that these rare special conditions will be met.
Technically, the model's state space is broken down into smaller
parts, and test case generation is performed for each subspace. The
smaller a state space, the more likely it is to reach its elements.

\subsubsection{Implementation coverage} 
\label{subsubsec:implcov}
Technical constraints with batch
processing made it impossible to run the same set of experiments on
the implementation (Sec.~\ref{subsubsec-errordetection}).
Because of the limited number of evaluated test suites we cannot
display the evaluation of coverage with an increasing number of test
cases. Instead, we display the relationship between model coverage and
implementation coverage (Fig.~\ref{fig:cov-vs-cov}) for test suites
with a fixed number of elements. These test suites form a superset of
those regarded in Fig.~\ref{fig:kinds-of-errors}; that not all of them
were considered in the error analysis is a consequence of the effort that is necessary to
assign failures to failure classes (Sec.~\ref{subsubsec-errordetection}).

That implementation coverage does not exceed 75\% is a result
of the abstractions applied to the model: we excluded most C functions
from the measurements that had no counterparts. However, as
mentioned above, some of the behavior abstracted in the model is
scattered through the code, and we did not touch these parts. One can
see that test suites that were built with
randomness ($\mathcal{B}$, $\mathcal{C}$, $\mathcal{D}$) yield rather different coverages in their own
classes.  This is likely due to random influences: as our
measurements and the 98\% confidence intervals in
Fig.~\ref{fig:cumulated-coverage} indicate at least for the model,
test suites from one category tend to yield rather constant coverages.

On average, the random suites $\mathcal{C}$ and $\mathcal{D}$ yield roughly the
same implementation coverage. As in the case of the model, coverage
tends to increase for suite $\mathcal{B}$.  There is a moderate
positive correlation between coverages
(correlation coefficient $r=.63$; $P\leq.001$). We
expected to see a stronger correlation on the grounds of the argument
that the ``main'' threads of functionality are identical in the
model and an implementation. This was not confirmed. The
figure suggests that there is a rather strong (the small number of
measurements forbids a statistical analysis) correlation of coverages
if only the manually derived  suites $\{\mathcal{E},\mathcal{F},\mathcal{G},\mathcal{A}\}$ are regarded.

While the manually built model-based test suite $\mathcal{A}$ yields higher model
coverage than the tests in $\mathcal{B}$---as explained above---it exhibits a
lower implementation coverage than $\mathcal{B}$. This, again, is a result of the
fact that the implementation ran into some branches that were not
modeled.

\subsubsection{Coverage vs. Error Detection}
\begin{figure}[t]
  \includegraphics[width=.46\textwidth]{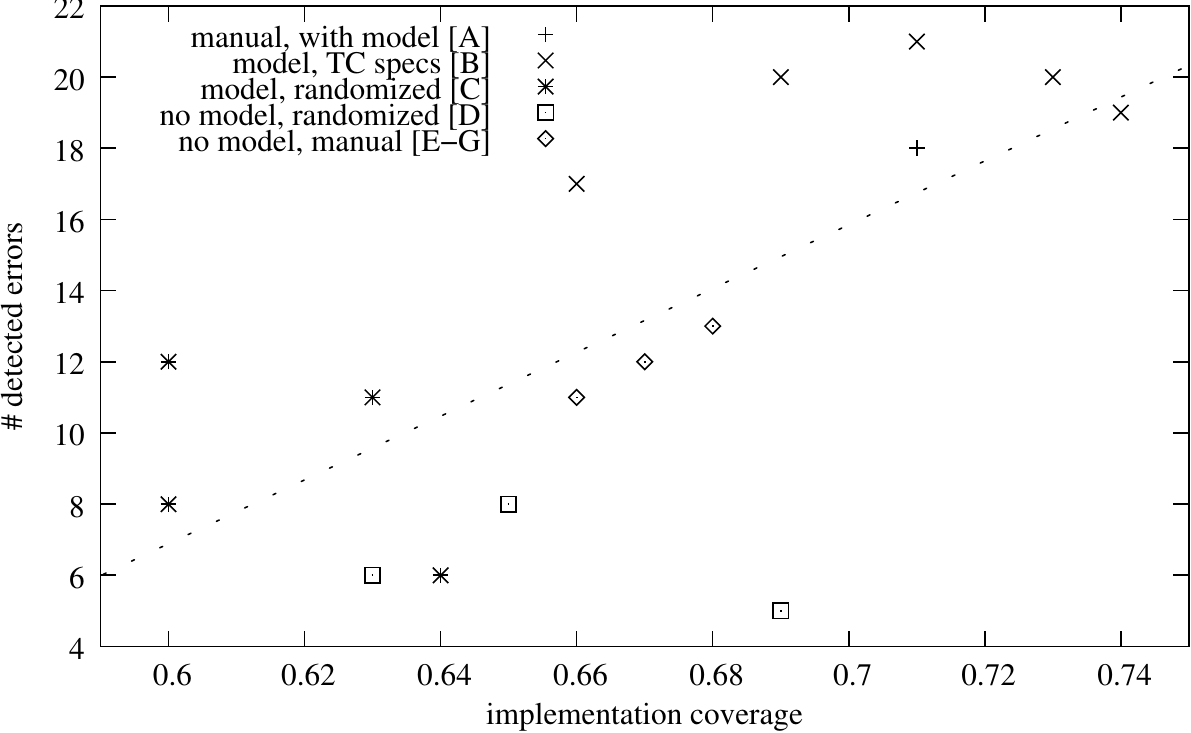}
  \caption{Implementation coverage vs. errors}
  \label{fig:implcov-vs-err}
\end{figure}

\begin{figure}[t]
  \includegraphics[width=.46\textwidth]{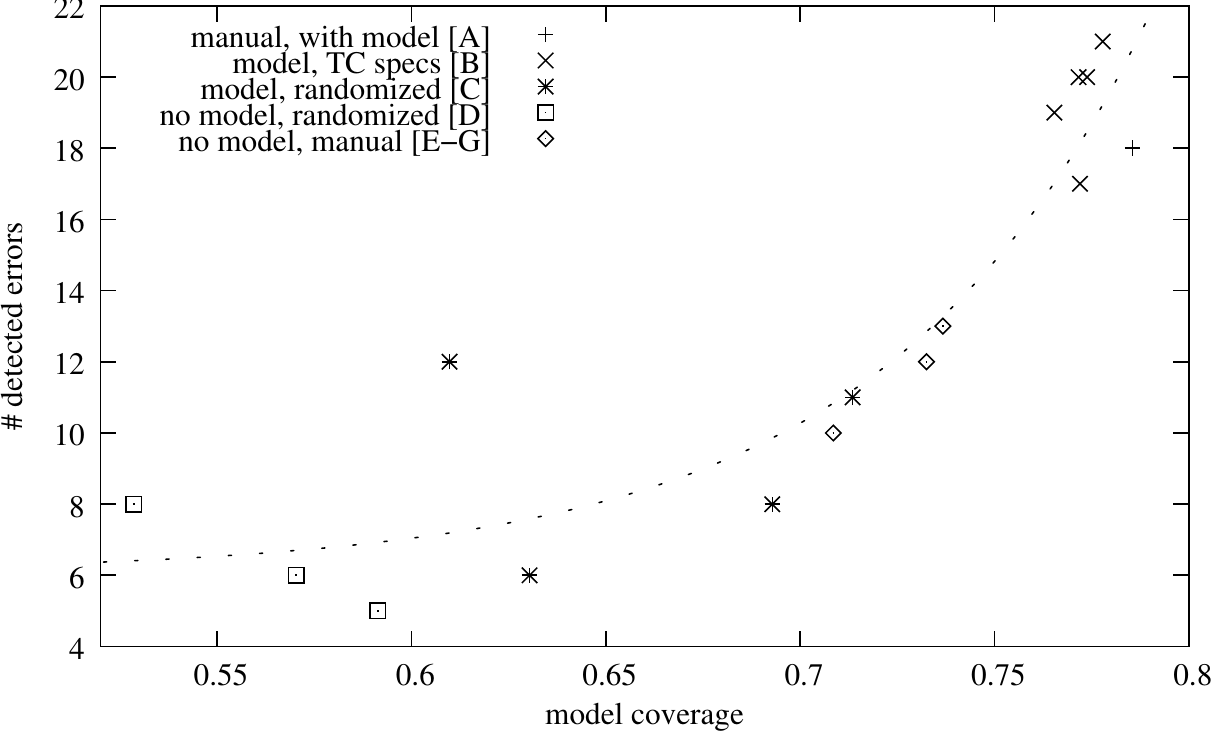}  
  \caption{Model coverage vs. errors}
  \label{fig:modcov-vs-err}
\end{figure}

A combination of the data from Secs.~\ref{subsubsec-errordetection}
to~\ref{subsubsec:implcov} is given in Figs.~\ref{fig:implcov-vs-err}
and~\ref{fig:modcov-vs-err}.
Both figures suggest a positive correlation between C/D coverage and
error detection. Data is more scattered in the case of implementation
coverage: correlation coefficient $r=.68$ ($P\leq.001$) for the
implementation. Correlation is $r=.84$ ($P\leq.0001$) for the model with a
logarithmically transformed ordinate.
We
observe in Fig.~\ref{fig:implcov-vs-err} that test suite $\mathcal{D}$ yields a
comparatively high coverage but finds few errors. As above, this is
explained by the fact that implementation coverage includes
functionality that is not implemented in the model, most importantly,
timing issues.  Furthermore, one can see that at high coverage levels,
increasing coverage does not necessarily increase the number of
detected errors.

\subsubsection{Summary}
As a bottom line, we observe the following.
\begin{enumerate}
\item The use of models significantly increases the number of detected
  requirements errors. Roughly, the number of detected programming
  errors does not depend on the use of a model.  Purely random tests
  $\{\mathcal{C},\mathcal{D}\}$, both with and without model, detect fewer errors than all
  other test suites.
\item None of the test suites detected all errors. When comparable
  numbers of tests are taken into account, hand-crafted model-based
  tests detect as many errors as automatically generated tests. When
  compared to hand-crafted model-based tests, six times (or even 9
  times if one subscribes to the argument of
  Sec.~\ref{subsubsec-errordetection}) more automatically generated
  tests detect three additional errors. That different test suites
  detect different errors suggests that a combination of test suites
  is preferable.
\item C/D coverages of model and implementation correlate moderately.
\item Overall, C/D coverage positively correlates with error
  detection, but higher coverage does not necessarily imply a higher
  error detection rate.
\item The rather high number of remaining requirements errors suggests
  that MSC-based requirements documents need to be complemented by the
  model itself.
\end{enumerate}

%%% Local Variables: 
%%% mode: latex
%%% TeX-master: "main"
%%% End: 

\section{Discussion}
\label{sec:discussion}

That the use of executable behavior models helps with clarifying
requirements and detecting errors does not surprise us: the behavior
model is an abstract prototype of the SUT. We consider it remarkable
yet not surprising that the number of detected \emph{programming}
errors is roughly independent of the use of models.

That the benefits of automation deserve some scrutiny corresponds to
our gut feeling of earlier studies
\cite{PhilippsPSAKS03,psak:MBTinhouse:04}.  ``Automation'' must be
taken with care. One, we still need humans to formulate test case
specifications; that structural criteria alone do not suffice as basis
for test case generation is widely undisputed.  (The use of test case
specifications also exhibits the intrinsic value of providing
rationales for test cases.)  Two, we had to perform some manual
optimizations in the generated CLP code: like all approaches to test
case generation we know of, our approach is not entirely a
push-button technology yet. 

Having said this, automation is indeed helpful when changes in the
model have to be taken into account. Provided that test case
generation is a push-button technology, it is obviously simpler to
automatically generate new tests than to hand-craft them. It is
possible to conceive and hand-craft 100 tests in a few hours, but this
becomes more complicated for 1,000
tests. 
Recall how a significant increase in automatically generated
model-based tests revealed some additional errors.  Obviously, the
length of the test cases---the number of steps that
must be performed---matters in a similar way. 
However, the number of test cases must be restricted to a minimum
because they not only have to be applied but also to be evaluated: if
there is a deviation in behaviors, then the test run must be manually
inspected. If 100 tests detect the same error, this becomes tedious.
In addition, in the case of (the software-in-the-loop simulation of)
the embedded system of our study, each test takes at least 10 seconds
because of hardware limitations. This naturally restricts the number
of tests that can be run. Furthermore, we found that purely randomly
generated tests are difficult to interpret because they correspond to
highly ``non-standard'' behavior.

Counting failures for reactive systems is non-trivial. 
When the behaviors of model and implementation differed at a certain
moment in time, they tended to differ for the rest of the test case,
too.  We tried to associate a maximum number of errors to a test run,
but were in doubt sometimes: in our statistics, the majority of test
cases revealed not more than one error.

It is difficult to draw conclusions from the moderate correlation
between model coverage and implementation coverage. Using coverage
criteria as test case specifications for automated test case
generation relies on their suspected ability to detect errors. In
addition to the ongoing controversy on this subject, our results
suggest some care
with directly transferring findings on \emph{implementation coverage}
to \emph{model coverage}. Model coverage, as we define it, is clearly
dependent on the simulation code generator that is used.

One must be careful to generalize. When comparing test suites built by
different teams, which is the case for our test suites $\mathcal{A}$
and $\mathcal{G}$, one must take into account the fact that different
people in different contexts with different knowledge of the system
conceived them (cf. Hamlet's comments \cite{hamlet:compareTesting:89},
and the findings of Hutchins et al.
\cite{hutchins:structCovEffective:94} that indicate that test suites
derived by different test engineers---or even different test suites
derived by the same engineer---vary w.r.t. effectiveness). While we
consider it possible to generalize our findings to other
event-discrete embedded devices with almost no ordered data types, we
cannot say whether the same is true for discrete-continuous embedded
systems or business information systems.  As mentioned above, it is,
in general, likely that the benefits of automation are greater if
significantly more tests could be run. This is not always the case for
embedded systems.

We are also aware that we used one specific modeling language, and
tested an implementation at a certain stage of development. We do not
know if our findings generalize for implementations in a more mature
state.  Furthermore, we cannot judge whether or not different coverage
criteria, particularly those based on data flow, exhibit the same
characteristics. In terms of test case generation technology, we do
not think that our approach is fundamentally different from others
(see Sec.~\ref{sec:relwork}).

%%% Local Variables: 
%%% mode: latex
%%% TeX-master: "main"
%%% TeX-master: "main"
%%% End: 

\section{Related Work}
\label{sec:relwork}

Test case generation on the grounds of structural criteria with model
checkers or symbolic execution has been proposed by different authors,
both for application to models of the implementation and to
environment models
\cite{hong:coverage:02,p:mcdctest:03,ammanblack:testingmcmutation:99,heimdahl:coverageMC:04,PhilippsPSAKS03}.
\autofocus{} models
can be subjected to (bounded) model checking, 
but this was not applicable with the current technology because of the
recursive data structures.
The model's complexity also inhibited successful application of our
own test generation technology for MC/DC \cite{p:mcdctest:03}. We
suspect
that even if we tightly restricted the recursive structures, 
a model checker couldn't cope with the model's complexity.

For a review of model-based test case generators, we refer to earlier
work \cite{psak:MBTinhouse:04}.

The present work uses coverage criteria to measure test cases, but not
to generate them.  Instead, we stick to a combination of using
functional test case specifications and random testing
\cite{duran:random:84,hamlet:random:90,gutjahr:random:99}---which,
sooner or later, is used in many test case generators, and which is
also induced by the search strategy of model checkers---but restrict
the sample space by means of test case specifications. In other words,
we randomly generate tests for ``slices'' of the model, and these
slices roughly correspond to the main modes of operation. In this
sense, we combine functional with random testing. This procedure
exhibits the advantage of yielding rationales for test cases.  The
test case specifications were written after the model was completed,
and we hence did not investigate their use in specification documents.
In conformance with intuition, among others, the studies by Heimdahl
and George \cite{heimdahlgeorge:testsuitereduction:04} and by Hutchins
et al. \cite{hutchins:structCovEffective:94} indicate that different
test suites with the same coverage may detect fundamentally different
numbers of errors.

Heimdahl et al. recently found that coverage-based tests generated by
symbolic model checkers must, in terms of failure detection, be
regarded with care \cite{heimdahl:coverageMC:04}.
Well-known studies
\cite{frankl:condcov:98,frankl:expCoverage:93,hutchins:structCovEffective:94,ntafos:coverageEffectiveness:84,girgis:coverageeffectiveness:86}
are concerned with the failure detection capabilities of coverage
criteria.  In sum, they are rather inconclusive.  Hutchins et al.
\cite{hutchins:structCovEffective:94} use test suites that were
manually generated on the grounds of the category partition method,
and then augmented in order to increase coverage.  The others use
randomly generated tests, and do not take into account human test
selection capabilities. All these studies do not study the
relationship between automatically and manually generated tests;
instead, the focus is on comparing tests that satisfy coverage
criteria on the grounds of control and data flows.  The studies of
Ntafos \cite{ntafos:coverageEffectiveness:84} and Hutchins et al.
\cite{hutchins:structCovEffective:94} are based on mutation testing or
fault seeding with the respective inherently critical assumptions.
Like Frankl et al.  \cite{frankl:condcov:98,frankl:expCoverage:93}, we
do not use mutation analysis for measuring effectiveness but, instead,
stick to actual errors. All these studies differ from ours in that
they are concerned with finding \emph{at least} one---or even the only
one---error.

There are few studies that investigate the relationship between
model/specification coverage and error detection---with notable
exceptions
\cite{heimdahl:coverageMC:04,weyuker:booleanSpecErrorDetect:94}. We
use coverage criteria at the level of generated simulation code rather
than at the specification level
\cite{offutt:speccriteria:99,weyuker:booleanSpecErrorDetect:94}
because we don't know of dedicated coverage criteria for EFSMs with
complex action languages: full-fledged recursive first-order
functional programs in transition guards and assignments.

Finally, Baresel et al. study the relationship between model and
implementation coverages \cite{baresel:modcovcodecov:03}: model
coverage is not defined by referring to generated code, and they find
dedicated model coverage criteria to correlate with classical coverage
criteria on generated code.

%%% Local Variables: 
%%% mode: latex
%%% TeX-master: "main"
%%% End: 

\section{Conclusions and Future Work}
\label{sec:concl}

Our study  substantiates earlier findings
that building  a prototype helps with improving requirements specifications.  The
use of models pays off when it comes to detecting failures by means of
model-based tests: two to six times more requirements errors could be
found. Recall that we used the model to update the specification
MSCs---some MSCs were corrected, seven were added. The rather high
number of remaining requirements errors in the implementation---issues
that were not clear enough in the specifications and that were not
captured by MSCs---suggests a need for complementing MSC-based
requirements documents. One could include the model itself into the
specification documents. This would require an additional overhead in
terms of documentation of the model. One could also include generated
tests, as MSCs, into the specification. However, there will always be
unspecified parts of the behavior, a consequence of the existential
nature of MSCs. A combination of both appears reasonable yet costly.

Programming errors are found more or less regardless of the use of a
model.  Automated test case generation did not yield more errors when
a comparable number of hand crafted model-based tests were applied.
However, we found that significantly more tests detect three additional
errors, or 11\%. We measured coverage at the levels both of the model
and the implementation. C/D coverages correlate moderately, likely a
consequence of abstractions in the model.  On the other hand, both
exhibit a positive correlation with error detection.  However,
increasing C/D coverage does not necessarily imply a larger number of
detected errors. This leads us to regard the use of (this) coverage
metrics with some skepticism.

In our context, automated test case generation refers to generating
tests---input and expected output---from a model and a set of
constraints that characterize ``interesting'' behaviors of the system.
These constraints were freely combined into test case specifications.
In this sense, we do functional testing with random selection. We did
not use structural criteria as test case specifications, and we think
that our results might stimulate further research in terms of
empirical investigations of the effectiveness of model-based test case
generation on the grounds of structural criteria only
\cite{heimdahl:coverageMC:04}.

We believe that the use of (behavior) models will become increasingly
popular in software/systems development. If automated test case
generation can, unlike model checking at present, be turned into a
push-button technology, it is a valuable add-on to hand-crafted tests:
if generating, running, and evaluating tests come at no cost, there is
no objection to using this technology. To the contrary, automated
tests detected errors that humans did not find.

Of course, generalizations must be applied with care. We have provided
several caveats and leave it to future studies to confirm or reject
the implications of our results.

%%% Local Variables: 
%%% mode: latex
%%% TeX-master: "main"
%%% End: 

Apart from the number and length of test cases, three major parameters
influence the effectiveness of automatically generated tests: adequacy
and level of detail of both model and test case specifications, and
adequacy of the generation technology itself. We have argued about
technology above. Like programming, building the model and choosing an
adequate level of abstraction is witchcraft at present. We consider
domain-specific modeling patterns as a promising step where the
restriction---to a product line, or a domain---remains to be
determined. In terms of test case specifications, for some application
areas like information security, large bodies of knowledge on
historical problems exist. Regardless of the domain, we believe that
turning such knowledge into libraries of explicit test case
specifications is likely to boost effectiveness of automated test case
generation.

We think that the general approach of testing on the grounds of
different levels of abstraction is also promising for mixed
discrete-continuous \cite{hahn.ea:prototype-based:2003} and real-time
systems, and we acknowledge the need for respective empirical
evaluations.

While the network controller is deterministic, many ideas of
model-based testing also apply to non-deterministic systems. We are
currently working on the generalization.

Empirical evaluations that generalize the present study are currently
organized. We plan to perform a study like this one with different
modeling languages and test case generators, and we are also re-doing
this study with different embedded systems. The investigation of
business information systems appears particularly interesting because
of a possibly higher number of tests that can be applied. Automated
regression testing immediately comes to mind.

Further planned studies are concerned with the efficiency of
model-based testing. This includes estimates on the impact of an
error's severity into the respective test case specifications.
Statistical user profiles could help identify the most common failures
\cite{musa:SWRE:04}.

The economics of using explicit behavior models in the development
process are not understood yet. In particular, it is not
clear if the life-cycle spanning synchronization of a model w.r.t.
possibly different implementations is economically efficient.

%%% Local Variables: 
%%% mode: latex
%%% TeX-master: "main"
%%% End: 

\emph{Acknowledgment.} Bernhard Seybold provided useful comments on a draft version of this paper.

%%% Local Variables: 
%%% mode: latex
%%% TeX-master: "main"
%%% End: 

%-------------------------------------------------------------------------

%-------------------------------------------------------------------------

%-------------------------------------------------------------------------
%\nocite{ex1,ex2}
\bibliographystyle{plain}
\bibliography{icse05}
\end{document}